\documentclass[fdp,a4paper,fleqn]{w-art}
\usepackage{times,cite,w-thm}
\usepackage{amsmath}
\usepackage{amsfonts} 
\usepackage{amssymb} 
\usepackage{amsthm}
\usepackage{graphicx}

\theoremstyle{plain}

\theoremstyle{definition}

\usepackage[]{graphicx}
\newcommand{\cT}{{\mathcal{T}}}
\newcommand{\cN}{{\mathcal{N}}}
\newcommand{\cF}{{\mathcal{F}}}
\newcommand{\cM}{{\mathcal{M}}}

\newcommand{\Tr}{\mathrm{Tr}\,}
\newcommand{\tr}{\mathrm{tr}\,}
\newcommand{\ft}[2]{{\textstyle\frac{#1}{#2}}}
\newcommand{\eq}[1]{Eq.~(\ref{#1})}
\newcommand{\ii}{\mathrm{i}}

\newcommand{\bone}{\mathbf{1}}
\newcommand{\diag}{{\rm diag}\,}
\newcommand{\R}{{\mathbb R}}

\newdimen\tableauside\tableauside=1.0ex
\newdimen\tableaurule\tableaurule=0.4pt
\newdimen\tableaustep

\def\phantomhrule#1{\hbox{\vbox to0pt{\hrule height\tableaurule
width#1\vss}}}
\def\phantomvrule#1{\vbox{\hbox to0pt{\vrule width\tableaurule
height#1\hss}}}
\def\sqr{\vbox{%
 \phantomhrule\tableaustep
\hbox{\phantomvrule\tableaustep\kern\tableaustep\phantomvrule\tableaustep}%
 \hbox{\vbox{\phantomhrule\tableauside}\kern-\tableaurule}}}
\def\squares#1{\hbox{\count0=#1\noindent\loop\sqr
 \advance\count0 by-1 \ifnum\count0>0\repeat}}
\def\tableau#1{\vcenter{\offinterlineskip
 \tableaustep=\tableauside\advance\tableaustep by-\tableaurule
 \kern\normallineskip\hbox
   {\kern\normallineskip\vbox
     {\gettableau#1 0 }%
    \kern\normallineskip\kern\tableaurule}%
 \kern\normallineskip\kern\tableaurule}}
\def\gettableau#1 {\ifnum#1=0\let\next=\null\else
 \squares{#1}\let\next=\gettableau\fi\next}
\newcommand{\Yfund}{\tableau{1}}
\newcommand{\Ysymm}{\tableau{2}}
\newcommand{\Yasymm}{\tableau{1 1}}
\newcommand{\reps}[4]{(\mathbf{#1},\mathbf{#2},\mathbf{#3},\mathbf{#4})}
\newcommand{\repst}[3]{(\mathbf{#1},\mathbf{#2},\mathbf{#3})}
\newcommand{\Smod}{S_{\rm mod}}

\begin{document}
\pagespan{1}{}

\title[Stringy instantons]{Stringy instantons and dualities}

\author[M. Frau]{Marialuisa Frau\inst{}%
  \footnote{\quad E-mail:~\textsf{frau@to.infn.it}}}
\address[\inst{}]{Dipartimento di Fisica Teorica, Universit\`a di Torino and I.N.F.N., sezione di Torino\\
via P. Giuria 1, I-10125 Torino (Italy)}
\begin{abstract}
We discuss non-perturbative corrections to the gauge kinetic functions in 
a four-dimensional $\mathcal{N} = 2$ gauge theory realized with a
system of D7/D3-branes in a compactification of 
type I$^\prime$ theory
on $\mathcal{T}_4/\mathbb{Z}_2\times
\mathbb{Z}_2$. The non-perturbative contributions arise when D(--1) branes, corresponding to stringy
instantons, are added to the system; such contributions can be explicitly evaluated using localization
techniques and precisely match the results predicted by the heterotic/type I$^\prime$ duality. This 
agreement represents a very non-trivial test of the stringy multi-instanton calculus.
\end{abstract}

\maketitle

\section{Introduction}

It has been recently found \cite{Blumenhagen:2009qh}
that certain classes of D-brane instantons 
arising in intersecting brane models can generate effective interactions 
at energies that are not linked to the gauge theory scale, and 
for this reason they are usually called ``stringy'' or ``exotic'' instantons.
This feature is very welcome in
the search of semi-realistic string scenarios for the physics beyond the Standard Model.
It is therefore of the greatest importance to devise techniques 
to determine quantitatively such exotic non-perturbative corrections through
their explicit realization at the string level. 
In this context both the usual gauge instantons and the exotic ones
can be obtained from Euclidean branes entirely wrapping some cycle of the internal space. 
Depending on whether this cycle coincides or not with the one wrapped by the space-filling
D-branes on which the gauge theory is defined, the Euclidean branes correspond 
to gauge or exotic instantons, respectively.

In the simplest cases, $4d$ gauge instantons can be realized with
bound states of space-filling D3-branes and point-like D(-1)-branes
\cite{Witten:1995gx}. In these systems
the massless sector of open strings having at least one endpoint on 
the D(-1)'s is in one-to-one correspondence with the moduli 
of the gauge instanton solution.
Actually, also the effective action on the moduli
space, the rules of the instanton calculus and the profile of the classical solution
can be explicitly obtained in this way 
\cite{Green:2000ke,Billo:2002hm}.

In the exotic cases, the gauge and instantonic branes intersect non-trivially in 
the internal space and thus the open strings stretching between
them have extra ``twisted'' directions and some instanton moduli 
(specifically those related to sizes and gauge orientations)
disappear from the spectrum. Their supersymmetric fermionic partners remain
massless, and when integrated out, they can 
lead to the effective interactions we alluded to above.
A very simple example of this phenomenon occurs in the D(-1)/D7 brane system,
which exhibits the world-sheet features of exotic instantons since
mixed open strings have eight twisted directions.
By adding O7-planes, this system can be embedded in type I$^\prime$ string theory,
a setup which possesses a computable perturbative heterotic dual.
The non-perturbative contributions of D-instantons to the effective action on
the D7-branes can be explicitly computed as integrals over the moduli space via
localization techniques, in strict analogy with what is done for usual gauge instantons
\cite{Nekrasov:2002qd}. One finds that all D-instanton
numbers correct the quartic gauge couplings of the $8d$ gauge
theory of the D7-branes\cite{Billo':2009gc,Billo:2009di,Fucito:2009rs}, 
and this whole series of terms matches the 
perturbative result obtained in the dual heterotic string theory.

In this contribution we briefly describe one example of exotic instanton
calculus in a $4d$ setup that has been developed in \cite{Billo':2010bd}. 
We consider a perturbatively conformal $\cN=2$ gauge theory that admits a 
brane realization where exotic instantons generate a
whole series of corrections to the quadratic gauge couplings and possesses 
a calculable heterotic dual against
which these corrections can be checked.
This provides a very non-trivial check of the 
correctness of this approach to the exotic instanton calculus. 


\section{A $\mathcal{N}=2$ conformal model from an orbifold of type I$^\prime$}
\label{sec:Imodel}

We consider a $\mathcal N=2$ orientifold compactification of
type IIB string theory on $\cT_4\times \cT_2$.
The action of the orientifold generators selects 4 O7-planes 
located at the invariant points of $\cT_2$ and 64 O3-planes located at the fixed points 
of $\cT_4\times \cT_2$.
The global cancellation of the RR tadpoles requires the
presence of 16 dynamical D7-branes transverse to $\cT_2$
and of 16 dynamical ``half`` D3-branes transverse to the internal 6-torus.
We choose to cancel \emph{locally} the RR charges in $\cT_2$ by
placing exactly 4 D7-branes and 4 half D3-branes on top of each O7-plane.
The D3's could then be distributed over the 16 orbifold fixed points 
that are common to a given O7-plane.
For sake of simplicity we place them at distinct points of $\cT_4$
and we focus on the gauge theory leaving in one of the O7 fixed plane.

The $4d$ field theory leaving on the D7 world-volume at the selected fixed 
plane is a conformal $\cN = 2$ $\mathrm{U}(4)$ SYM theory containing
one adjoint vector multiplet, two antisymmetric hypermultiplets and 
four fundamental hypermultiplets which are charged under a $\mathrm{U}(1)^4$
flavour group.
The quadratic effective action for the gauge fields can be described by 
holomorphic Wilsonian couplings \cite{Dixon:1990pc}
that have the following structure:
\begin{equation}
f = f_{(0)} + f_{(1)}+f_{\mathrm{n. p.}}
\label{f01}
\end{equation}
where the subscripts $_{(0)}$ and $_{(1)}$ refer
to the tree-level and 1-loop contributions, while the last term accounts for
possibile non perturbative corrections. 
Writing the effective action in terms of the $\mathcal{N}=2$ multiplet encoding 
the $\mathrm{U}(4)$ gauge degrees of freedom,
$\Phi(x,\theta) = \phi(x) + \theta^\alpha \Lambda_\alpha(x) + 
(\theta \gamma^{\mu\nu} \theta)\, F_{\mu\nu}(x)$,
we see that there are two possible colour structures, 
each with its own coupling:
\begin{equation}
\label{action}
S = \int d^4x d^4\theta \left\{f\, \Tr \Phi^2 + f'(\Tr\Phi)^2\right\} +\mathrm{c.c}~.
\end{equation}
The tree-level value for the single trace coupling $f$ can be deduced from the 
Born-Infeld action and is
\begin{equation}
f_{(0)}= -{\rm{i}}\,t \ \ \ \ \ \ \ \ \ \ \
({\rm Re}\,t=\frac{\theta_{
YM}}{2\pi}~,~~
{\rm Im}\,t=\frac{4\pi}{g_{
YM}^2}
\sim \frac{Vol({\mathcal T}_4)}{g_s}) ~;
\label{f0}
\end{equation}
on the other hand the tree level value for the double trace coupling
$f'$ is vanishing. 

The only perturbative corrections to $f$ and $f'$ come from the 1-loop 
threshold corrections, which in turn are related in an universal way to 
the string 1-loop two point-functions.
The correction to the single trace coupling $f_{(1)}$ is expected to vanish, 
since the gauge theory is conformal, and in fact  
the 1-loop string diagrams that contribute to the single trace coupling add
up to zero.
On the contrary, the 1-loop diagrams that contribute to the double trace structure
give a non vanishing result, due to the massless states winding on ${\mathcal T}_2$, and we have 
\begin{equation}
f'_{(1)} = -8 \log \eta(U) ~,
\label{fIaa}
\end{equation}
where $U$ is complex structure of ${\mathcal T}_2$.

In the next sections we will study the non-perturbative
corrections $f_{\mathrm{n. p.}}$ and $f'_{\mathrm{n. p.}}$ induced by D-instantons.


\section{D-instantons and their moduli spectrum}
\label{sec:Dmod}

The orientifold projection that defines our model is 
compatible both with  Euclidean E3-branes wrapped on $\cT_4$
that represent ordinary gauge instantons for the field
theory living on the D7-branes, and 
D-instantons which describe truly stringy instanton configurations for
the D7-brane gauge theory \cite{Billo':2009gc,Billo:2009di,Fucito:2009rs}.
Here we only discuss the contributions produced by the D(-1)-branes,
and show that they correct non-perturbatively the gauge kinetic functions
of the $\mathcal N=2$ $\mathrm{U}(4)$ theory discussed in Sect.~\ref{sec:Imodel}.

Again we focus on the four D7-branes located at one of the orientifold fixed points, 
and place on them a number of fractional D-instantons. However, since there are also
four D3-branes distributed in four different orbifold fixed points, we have to
distinguish between two possibilities, depending on whether the D-instantons are
at an empty fixed point or occupy the same position of one of the D3-branes. 
In the first case (case \emph{a)})
only the (-1)/(-1) and (-1)/7 open strings support massless moduli,
because the (-1)/3 strings have always a non-vanishing stretching energy
due to the separation between their endpoints.
In the second case (case \emph{b)})
we can find massless excitations 
also in the spectrum of the (-1)/3 strings.
Consistently with the orientifold projections, when we set 
$k$ ``half'' D-instantons at a given fixed point,
the instanton moduli
organize in representation of $\mathrm{U}(k)$ and in representations of the
Lorentz symmetry group, which in our local system is broken to
$\mathrm{SO}(4)\times \widehat{\mathrm{SO}}(4) \times \mathrm{SO}(2) = 
\mathrm{SU}(2)_+\times \mathrm{SU}(2)_- \times 
\widehat{\mathrm{SU}}(2)_+\times \widehat{\mathrm{SU}}(2)_- \times
\mathrm{SO}(2)$.

The (-1)/(-1) moduli form the so-called neutral 
moduli sector, since they do not transform under the $\mathrm{U}(4)$ gauge group and are 
common to both case \emph{a)} and \emph{b)}. 
They comprise four complex scalars 
transforming as vectors of the $\mathrm{SO}(4) \times \widehat{\mathrm{SO}}(4)$ groups 
that rotate the coordinates of the D7 world volume,
a complex scalar $\chi$ and their fermionic partners.
The charged moduli sector accounts for the (-1)/7 open strings. 
Since there are eight directions with mixed boundary conditions
we only find a physical fermionic modulus $\mu'$.
Finally, the flavored sector of the instanton moduli space arises from the 
(-1)/3 open strings. 
In our model, this sector exists only in case \emph{b)}, when the D(-1)'s 
and the D3's occupy the same fixed point.
It will be useful however to consider the generalized case with $m$
half D3-branes supporting a $\mathrm{U}(m)$ symmetry, so that the configuration \emph{a)} 
corresponds to $m=0$ and the configuration \emph{b)} corresponds to $m=1$.
In this case, one finds two complex variables transforming
as chiral spinors with respect to $\mathrm{SO}(4)$, and two spinors
of opposite chiralities with respect to $\widehat{\mathrm{SO}}(4)$.
All physical moduli and their transformation properties are summarized in Tab.~1.
\begin{small}
\begin{table}[htb]
\begin{center}
\begin{tabular}{|c|c|c|c|c|c|
c|
}
\hline
\begin{small} {\phantom{\vdots}} neutral\;\;\;\end{small}
&\begin{small} $\mathrm{SU}(2)^4$ \end{small}
&\begin{small}$\;\;\;\;\;\mathrm{U}(k)\;\;\;\;\;$ \end{small}
& \!\!\!\!\!\!\!\!{}
&\begin{small} {\phantom{\vdots}}charged\end{small}
&\begin{small} $\mathrm{SU}(2)^4$ \end{small}
&\begin{small} $\!\!\!\!\!\!\mathrm{U}(k)\times \mathrm{U}(4)\!\!\!\!\!\!$ \end{small} 
\\
\hline
{\phantom{\vdots}}
$\begin{array}{c} B_\ell \\  M_{\dot \alpha  a} \end{array}$
& $\begin{array}{c}\reps 2211 \\ \reps 1221\end{array}$
& adjoint  
& \!\!\!\!\!\!\!\!{}
&{\phantom{\vdots}}$\begin{array}{c}\mu'\end{array}$  
& $\reps 1111$ 
& $(\Yfund,\overline{\Yfund})$ 
\\
\hline
{\phantom{\vdots}}
$\begin{array}{c} N_{\dot \alpha \dot a}
\end{array}$
& $\begin{array}{c}\reps 1212 
\end{array}$
& $\Yasymm+ \overline{\Yasymm}$ 
& \!\!\!\!\!\!\!\!{}
&\begin{small} {\phantom{\vdots}}flavoured \end{small}
&\begin{small} $\mathrm{SU}(2)^4$ \end{small}
&\begin{small} $\mathrm{U}(k)\times \mathrm{U}(m)$ \end{small} 
\\
\hline
{\phantom{\vdots}}
$\begin{array}{c} B_{\dot\ell} \\ M_{\alpha  \dot a} \end{array}$
& $\begin{array}{c}\reps 1122 \\ \reps 2112\end{array}$
& $\Ysymm+ \overline{\Ysymm}$ 
& \!\!\!\!\!\!\!\!{}
&$\begin{array}{c} w_\alpha \\ \mu_a \end{array}$
& $\begin{array}{c} \reps 2111 \\ \reps 1121 \end{array}$
& ($\Yfund,\overline{\Yfund})$ 
\\
\hline
{\phantom{\vdots}}
$\begin{array}{c}   N_{\alpha a} \\
\bar \chi \end{array}$
& $\begin{array}{c}\reps 2121 
\\ \reps 1111\end{array}$
& adjoint 
& \!\!\!\!\!\!\!\!{}
& $\begin{array}{c} \mu_{\dot a} 
\end{array}$
& $\reps 1112 $
& $(\Yfund,\Yfund)$ 
\\
\hline
\end{tabular}
\label{tab:2}
\caption{Spectrum of physical instanton moduli.}
\end{center}
\end{table}
\end{small}

Note that in addition to the physical moduli we have to consider
extra auxiliary fields, $d_{m}$, $D_{\dot \alpha \dot a}$,
$h_a$ and $h'$,
that linearize the quartic interactions among the moduli
and whose equations of motion generalize the 
ADHM constraints on the ordinary instanton moduli space.

\section{Non-perturbative corrections from localization formul\ae}
\label{sec:loc}
The corrections induced by D-instantons can be encoded in a 
non perturbative prepotential $\cF_{\mathrm{n.p.}}(\Phi)$,
which, taking into account the different instanton configurations 
and their multiplicity, can be written as
\begin{equation}
\label{prepsum}
 \cF_{\mathrm{n.p.}}(\Phi) = 12\, \cF^{(m=0)}(\Phi) 
 + 4\, \cF^{(m=1)}(\Phi)~.
\end{equation}

The prepotentials $\cF^{(m)}(\Phi)$ can be expressed as integrals 
over the ``centered'' moduli space 
(containing all moduli except the ``center of mass'' 
coordinates $x$ and $\theta$) of the instantonic branes.
To compute $\cF_{\mathrm{n.p.}}(\Phi)$ we exploit the fact 
\cite{Nekrasov:2002qd} that, after suitable deformations of the instanton action, 
the modular integrals localize around isolated points in the instanton moduli space.
To obtain explicit formulas we first take $\Phi=\diag(a_1,\ldots, a_4,-a_1,\ldots, -a_4)$,
where $a_u$ are constant
expectation values along the Cartan directions of $\mathrm{U}(4)$, and then
consider the $\epsilon$-deformed instanton partition function
\begin{equation}
\label{ipf0}
Z^{(m)}(a,\epsilon) =\sum_k q^k \,Z^{(m)}_k(a,\epsilon) 
= \sum_k q^k \int \!d\cM_{k,m} ~{\rm e}^{-\Smod^{\epsilon}(\cM_{k,m},a)}~.
\end{equation}
where $\Smod^{\epsilon}$ is obtained by deforming the moduli action
with Lorentz breaking terms parameterized by
four parameters $\epsilon_I$ describing rotations along the four Cartan
directions of 
$\mathrm{SO}(4) \times \widehat{\mathrm{SO}}(4)$.
{From} the string perspective, these deformations can
be obtained by switching on suitable RR background fluxes on the D7-branes, as
shown in \cite{Billo:2006jm,Billo:2009di,Billo':2008sp}.
Notice that integrals in (\ref{ipf0}) run over all moduli, including 
$x$ and $\theta$. In presence of the $\epsilon$-deformations it is rather easy
to see 
that the integration over the super-space yields a volume factor growing as
$1/(\epsilon_1\epsilon_2)$ in the limit of small $\epsilon_{1,2}$.  
Therefore, to obtain the integral over the centered moduli this factor has to be
removed. 
In addition, we have to notice that the $k$-th order in the
$q$-expansion receives contributions not only from genuine $k$-instanton
configurations but also from disconnected ones.
Thus, we are led to consider
\begin{equation}
\label{ipf}
\cF^{(m)}(a,\epsilon) =\epsilon_1 \epsilon_2  \log  Z^{(m)}(a,\epsilon) ~.
\end{equation}
The prepotential will be extracted from  $\cF^{(m)}(a,\epsilon)$ 
by sending
$\epsilon_I \to 0$ and $a \to \Phi$.

The localization procedure is based on the co-homological structure of the
instanton moduli action which is exact with respect to a suitable BRST
charge $Q$, namely
\begin{equation} 
\Smod= Q \Xi~.
\label{q}
\end{equation}

We can choose as $Q$ any component of the supersymmetry charges
preserved on the brane system. Of course, since these charges
transform as spinors of $\mathrm{SO}(4) \times \widehat{\mathrm{SO}}(4)$,
the choice of $Q$ breaks 
this symmetry to the $\mathrm{SU}(2)^3
\equiv
\mathrm{SU}(2)_- \times \widehat{\mathrm{SU}}(2)_- \times 
\diag\left[\mathrm{SU}(2)_+ \times
\widehat{\mathrm{SU}}(2)_+\right]
$ 
subgroup which preserves this spinor.
After this identification is made we can see that all the moduli 
but $\chi$ form BRST doublets, which we will schematically denote as 
$\big(\phi,\psi\equiv Q\phi\big)$,
and the moduli action can indeed be written in the form (\ref{q}).

To localize the integral over moduli space, it is necessary to make the charge
$Q$ equivariant with respect to all symmetries, which in our case are the gauge
symmetry $\mathrm{U}(k)\times \mathrm{U}(4)\times \mathrm{U}(m)$ and the
residual Lorentz symmetry $\mathrm{SU}(2)^3$. 
After the equivariant deformation, the charge $Q$ becomes
nilpotent up to an element of the symmetry group.  
In the basis provided by the weights
$\vec q \equiv \bigl({\vec q}_{\mathrm{U}(k)}, \vec q_{\mathrm{U}(4)},
{\vec q}_{\mathrm{U}(m)}, {\vec q}_{\mathrm{SU}(2)^3} \bigr)$,
$Q$ acts diagonally
\begin{equation}
\label{brspair3}
Q\phi_q = \psi_q~,~~~
Q\psi_q = \Omega_q \phi_q~,
\end{equation}
where
$\Omega_q = \vec\chi\cdot {\vec q}_{\mathrm{U}(k)} + \vec a \cdot 
 \vec q_{\mathrm{U}(4)} + \vec b \cdot {\vec q}_{\mathrm{U}(m)}
 + \vec \epsilon \cdot {\vec q}_{\mathrm{SU}(2)^3}$, 
parametrize the equivariant deformation
in terms of the Cartan components of the group parameters
$\vec \chi$, $\vec{b}$, $\vec{a}$ and $\vec \epsilon$. 
\footnote{The Cartan directions of the residual Lorentz group
$\mathrm{SU}(2)^3$ are parametrized by $\epsilon_I$ ($I=1,\ldots,4$) 
subject to the constraint
$\epsilon_1+ \epsilon_2+ \epsilon_3 +\epsilon_4=0$.}
\begin{table}[htb]
\begin{center}
\begin{tabular}{|c|c|c|c|
}
\hline
\begin{small} $(\phi,\psi)$ \end{small}
&\begin{small} $\mathrm{U}(k) \times \mathrm{U}(4)\times \mathrm{U}(m)$
\end{small} 
&\begin{small} $\mathrm{SU}(2)^3$ \end{small}
&\begin{small}  $\vec\epsilon\cdot \vec{q}_{SU(2)^3}$ \end{small}
\\
\hline
$(B_\ell,M_\ell)$ 
& $\bigl(\mbox{adj}, \bone, \bone\bigr)$ 
& $\repst 212$ 
& $ \epsilon_1,\epsilon_2$
\\
$(B_{\dot\ell},M_{\dot\ell})$ 
& $\bigl(\Ysymm, \bone, \bone\bigr) + \mbox{h.c.}$
& $\repst 122$   
& $ \epsilon_3,\epsilon_4$
\\
$(N_{\dot\alpha\dot a},D_{\dot\alpha\dot a})$ 
& $\bigl(\Yasymm, \bone, \bone\bigr) + \mbox{h.c.}$
& $\repst 221$   
& $ \epsilon_2+\epsilon_3,\epsilon_1+\epsilon_3$
\\
$(N_{m},d_{m})$ 
& $\bigl(\mbox{adj}, \bone, \bone\bigr)$ 
& $\repst 113$ 
& $ 0_{\R},\epsilon_1+\epsilon_2$
\\
$(\bar\chi,\eta)$ 
& $\bigl(\mbox{adj}, \bone, \bone\bigr)$ 
& $\repst 111$ 
& $ 0_{\R} $
\\
$(\mu',h')$ 
& $\bigl(\Yfund, \overline{\Yfund}, \bone\bigr) + \mbox{h.c.}$
& $\repst 111$    
& $ 0$
\\
$(w_\alpha,\mu_\alpha)$ 
& $\bigl(\Yfund, \bone, \overline{\Yfund}\bigr) + \mbox{h.c.}$
& $\repst 112$   
& $ 
(\epsilon_1+\epsilon_2 )/2$  
\\
$(\mu_{\dot a},h_{\dot a})$ 
& $\bigl(\Yfund, \bone, \Yfund\bigr) + \mbox{h.c.}$
& $\repst 121$ 
& $ 
(\epsilon_3-\epsilon_4 )/2$ 
\\
\hline
\end{tabular}
\caption{BRST structure and symmetry properties of the D(-1)/D3/D7 moduli.
The first two columns report the transformation properties under the
symmetry groups.
The last column collects the eigenvalues  $\vec\epsilon\cdot \vec{q}_{SU(2)^3}$
for the positive weights $\vec{q}$'s specified in the third column.}
\end{center}
\label{tab:bs}
\end{table}

After the complete localization the integral 
is given by  the (super)-determinant of $Q^2$ evaluated at 
the fixed points of $Q$ 
\cite{Nekrasov:2002qd,Bruzzo:2002xf},
and its explicit expression can be deduced by considering, for each
modulus $\phi$ in Tab.~2, the set of weights corresponding to its symmetry
representation. The explicit result is
\begin{eqnarray}
Z_k^{(m)}(a,b,\epsilon) &=& 
\left( \frac{s_3}{\epsilon_1 \epsilon_2 }
\right)^k\int \prod_{i=1}^k \!\frac{d{\chi_i}}{2\pi\ii} ~
\prod_{i<j}^{k} \big(\chi_i -\chi_j\big)^2\,\Big( (\chi_i -\chi_j)^2-s_{3}^2
\Big)\,
\nonumber\\
&&
\times  \prod_{i<j}^{k}\, \prod_{\ell=1}^{2} 
\frac{ \Big((\chi_i +\chi_j)^2-s^2_{\ell}\Big)}
{\Big((\chi_i-\chi_j)^2-\epsilon_{\ell}^2\Big)
\Big( (\chi_i+\chi_j)^2-\epsilon_{\ell+2}^2\Big)}\label{Z}
\\
&&\times 
\, \prod_{i=1}^k\left[\,\prod_{\ell=1}^{2} \frac{1}{\Big(4\chi_i^2
-\epsilon^2_{\ell+2}\Big)} \,
\prod_{r=1}^{m}
\frac{\Big(( \chi_i
+b_r)^2-\frac{(\epsilon_3-\epsilon_4)^2}{4}\Big)}{\Big((\chi_i
-b_r)^2-\frac{(\epsilon_1+\epsilon_2)^2}{4}\Big)} 
\,\prod_{u=1}^{n} \Big(\chi_i-a_u\Big)
\right]~.
\nonumber
\end{eqnarray}
The integral over  $\chi_i$ in this expression has to be thought of as a
multiple contour integral, according to the prescription introduced in
Ref.~\cite{Moore:1998et}.

In order to obtain the non-perturbative prepotential 
from the partition function $Z^{(m)}(a,b,\epsilon)$, we 
set $b_r=0$, since the D3-branes are fixed at one of the orbifold 
fixed-points and
we take the limit $\epsilon_I\to 0$ to remove the Lorentz breaking deformations. 
A simple inspection of the explicit results for $\log Z^{(m)}(a,\epsilon)$
\cite{Billo':2010bd} shows that this expression diverges as 
$1/(\epsilon_1 \epsilon_2\epsilon_3 \epsilon_4)$ in this limit. Such a
divergence is typical of interactions in eight dimensions, where the 
$\cN=2$ super-space volume grows
like $\int d^8x d^8 \theta \sim 1/(\epsilon_1 \epsilon_2\epsilon_3 \epsilon_4)$.
These contributions can be thought of as coming from regular
D(--1)-instantons moving in the full eight-dimensional world-volume of the
D7-branes and can in fact be associated to a universal quartic prepotential 
${\mathcal F}_{\mathrm{IV}}(a)$ \cite{Billo:2009di}
defined as
 \begin{equation}
 \label{F4}
 {\mathcal F}_{\mathrm{IV}}(a)
=  \lim_{\epsilon_I\to 0} \epsilon_1 \epsilon_2\epsilon_3 \epsilon_4  
 \log Z^{(m)}(a,\epsilon)~.
 \end{equation}
We can then extract a finite quadratic prepotential by subtracting the divergence coming from 
${\mathcal F}_{\mathrm{IV}}(a)$:
\begin{equation}
\label{F2}
{\mathcal F}_{\mathrm{II}}^{(m)}(a)
=  \lim_{\epsilon_{I}\to 0} \Big(
\epsilon_1 \epsilon_2 \log Z^{(m)}(a,\epsilon) 
-\frac{1}{\epsilon_3 \epsilon_4} {\cal F}_{\mathrm{IV}}(a) \Big)~.
\end{equation}
Since the moduli measure is dimensionless
no dynamically generated scale may appear and the contributions at \emph{all} 
instanton numbers must be constructed only out of the $a$'s; we find
\begin{equation}
\begin{aligned}
 \!\!{\mathcal F}_{\mathrm{II}}^{(m=0)}(a) 
&=\Big(\!\!-\sum_{i<j}a_ia_j \Big) \, q +
\Big(\sum_{i<j}a_ia_j-\frac14\,\sum_ia_i^2 
\Big)\,q^2 +\Big(\!\!-\frac{4}{3}\sum_{i<j}a_ia_j \Big)\,q^3+\cdots~,\\
\!\!{\mathcal F}_{\mathrm{II}}^{(m=1)}(a) 
&=\Big(3\sum_{i<j}a_ia_j \Big) \, q +
\Big(\sum_{i<j}a_ia_j+\frac74\,\sum_ia_i^2 
\Big)\,q^2 +\Big(4\sum_{i<j}a_ia_j \Big)\,q^3+\cdots~.
\end{aligned}
\label{F2b}
\end{equation}

We can now promote the vacuum expectation values $a$'s to the dynamical 
superfield $\Phi(x,\theta)$ and determine 
${\mathcal F}_{\mathrm{n.p.}}(\Phi)$ taking into account the contributions 
from the various $m=0,1$ configurations according to \eq{prepsum}.
Performing the $\theta$-integration, we then obtain the quadratic non-perturbative
action:
\begin{equation}
\label{snpq}
S_{\mathrm{n.p.}}= 4 \int d^4x\, \Big[ 2\big(\tr F\big)^2 - \tr F^2 \Big]
\,q^2 + O(q^4)~~+~
\mathrm{c.c.}
\end{equation}
and read the non-perturbative part of the holomorphic couplings.
Considering also the perturbative contributions written above we have
\begin{equation}
\label{concl_ffpI}
 f = -\ii t - 4\,q^2 + O(q^4)\phantom{\vdots}~,\ \ \ \ \ \ \ \ \ \ 
 f'= -8\log \eta(U)^2 +8 \,q^2 + O(q^4)\phantom{\vdots}~.  
\end{equation}

We would like to stress that the vanishing of the contributions at the one
and three instanton level is due to the non-trivial cancellations between
contributions coming from 
configurations \emph{a)} and \emph{b)}.

The heterotic model dual to the Type I$^\prime$ description of the previous sections
can be built from the 
U(16) compactification of the SO(32) heterotic string on $\cT_4/\mathbb Z_2$ 
(with standard embedding of the orbifold curvature into the gauge bundle)
and further reduced on $\cT_2$ with Wilson lines that break U(16) to
$\mathrm{U}(4)^4$.
The gauge kinetic terms in this heterotic set-up are corrected at 1-loop by an
infinite tower of world-sheet instantons wrapping $\cT_2$, which are dual to the
D-instantons of the type I$^\prime$ theory \cite{Bachas:1997mc}
and read \cite{Billo':2010bd}:
\begin{equation}
\label{concl_ffphet}
 f  = -\ii S + 8 \log\Bigg(\frac{\eta(\ft{T}{4})^2}{\eta
(\ft{T}{2})^2}\Bigg)~,\ \ \ \ \ \ \ \ \ \ 
 f'= -8\log \eta(U)^2 + 8 \log\Bigg(\frac{\eta(\ft{T}{2})^2}{\eta
(\ft{T}{4})^4}\Bigg)~. 
\end{equation}
These couplings are exact and
do not receive any kind of corrections beyond 1-loop. Therefore
they must contain all information, both
perturbative and non-perturbative, on the corresponding type I$^\prime$
couplings, including the (exotic) instanton corrections computed
above. Indeed, when we expand
for large values of $T$ and use the duality map that relate
the K\"ahler modulus of the heterotic theory $T$ to the axio-dilaton 
$\lambda$ of the type I$^\prime$ model: ${T}/{4} \longleftrightarrow \lambda$,
these heterotic formulas predict no instanton corrections at $k=1$ and $k=3$, 
and a relative coefficient $-2$ between the $k=2$ corrections to $f$ and $f'$, 
in perfect agreement with the results obtained in the type I$^\prime$ setting.

We regard these results as a nice and non-trivial confirmation of the validity
of the 
exotic instanton calculus, which can then be applied with confidence also to
four-dimensional theories and to models for which the heterotic dual is not
known or does not exist.

\end{document}